\documentclass{INTERSPEECH2023}

\interspeechcameraready
\usepackage{multirow}
\usepackage{amsfonts}
\usepackage{graphicx}
\title{An Enhanced Res2Net with Local and Global Feature Fusion for Speaker Verification}
\name{Yafeng Chen$^1$, Siqi Zheng$^1$, Hui Wang$^1$, Luyao Cheng$^1$, Qian Chen$^1$, Jiajun Qi$^2$}
\address{$^1$Speech Lab, Alibaba Group, Hangzhou, China \\
$^2$University of Science and Technology of China, Hefei, China}
\email{\{chenyafeng.cyf, zsq174630\}@alibaba-inc.com}

\begin{document}

\maketitle
 
\begin{abstract}
Effective fusion of multi-scale features is crucial for improving speaker verification performance. While most existing methods aggregate multi-scale features in a layer-wise manner via simple operations, such as summation or concatenation. This paper proposes a novel architecture called Enhanced Res2Net (ERes2Net), which incorporates both local and global feature fusion techniques to improve the performance. The local feature fusion (LFF) fuses the features within one single residual block to extract the local signal. The global feature fusion (GFF) takes acoustic features of different scales as input to aggregate global signal. To facilitate effective feature fusion in both LFF and GFF, an attentional feature fusion module is employed in the ERes2Net architecture, replacing summation or concatenation operations. A range of experiments conducted on the VoxCeleb datasets demonstrate the superiority of the ERes2Net in speaker verification. Code has been made publicly available at https://github.com/alibaba-damo-academy/3D-Speaker.

\end{abstract}

\section{Introduction}

Speaker verification (SV) is the task of determining whether a given speech utterance belongs to a claimed speaker identity. Traditional SV methods typically consist of two stages: a front-end \cite{DBLP:journals/csl/CampbellCRST06, DBLP:journals/taslp/KennyBOD07, DBLP:journals/taslp/DehakKDDO11} that generates fixed-dimensional speaker representations from the input signal, and a back-end such as probabilistic linear discriminant analysis (PLDA) \cite{DBLP:conf/iccv/PrinceE07} that compares the similarity between two speaker representations to determine whether they belong to the same speaker.

With the great success of deep learning, the vast majority of work has been focused on how to learn discriminative speaker embeddings through deep neural networks (DNNs). Most of them use convolutional neural network such as Residual Network (ResNet) \cite{DBLP:conf/cvpr/HeZRS16}, time-delay neural network (TDNN) such as ECAPA \cite{DBLP:conf/interspeech/DesplanquesTD20} and its variants \cite{DBLP:conf/interspeech/ChenGG21, DBLP:conf/slt/ZhouZW21, DBLP:conf/interspeech/LiuCWWHQ22, DBLP:journals/taslp/GuGZ23, DBLP:journals/corr/abs-2211-04168}. In ResNet-based models, short skip connections are used to fuse identity mapping features and residual learning features, allowing the model to more effectively propagate information through the network and improve the discrimination of the speaker embeddings. Meanwhile, ECAPA-based models utilize a multi-layer feature aggregation technique that aggregates shallow and deep features via summation, enabling the model to capture global patterns in the input signal.
So, whether explicit or implicit, intentional or unintentional, feature fusion is omnipresent for modern network architecture.  

However, a great variety of feature fusion methods aggregate the features in a layer-wise manner via element-wise addition or concatenation \cite{DBLP:conf/icassp/TangDHHZ19, DBLP:conf/interspeech/SeoRLLPOKK19, DBLP:conf/interspeech/HajaviE19, DBLP:conf/interspeech/JungKCJK20, DBLP:conf/interspeech/ZhangLWZH0LM22}. June et al. \cite{DBLP:conf/icassp/TangDHHZ19} proposed a multi-level pooling strategy to collect speaker information from both TDNN and long short-term memory (LSTM) layers. Acoustic features extracted from TDNN layer and LSTM layer are pooled and concatenated to obtain a robust speaker embedding. 
Seo et al. \cite{DBLP:conf/interspeech/SeoRLLPOKK19} proposed concatenating the pooled activation outputs from bottleneck architecture by shortcut connections to speaker embeddings. 
Hajavi et al. \cite{DBLP:conf/interspeech/HajaviE19} used weighted summation on pooled embeddings from multiple stages for aggregation. Jung et al. \cite{DBLP:conf/interspeech/JungKCJK20} proposed a top-down architecture with lateral connections to generate features with rich speaker information at all selected layers. Zhang et al. \cite{DBLP:conf/interspeech/ZhangLWZH0LM22} presented MFA-Conformer to capture global and local features effectively. The output feature maps from all conformer blocks are concatenated to aggregate multi-scale representations before final pooling. 

\begin{figure}[t]
  \centering
  \includegraphics[scale=0.7]{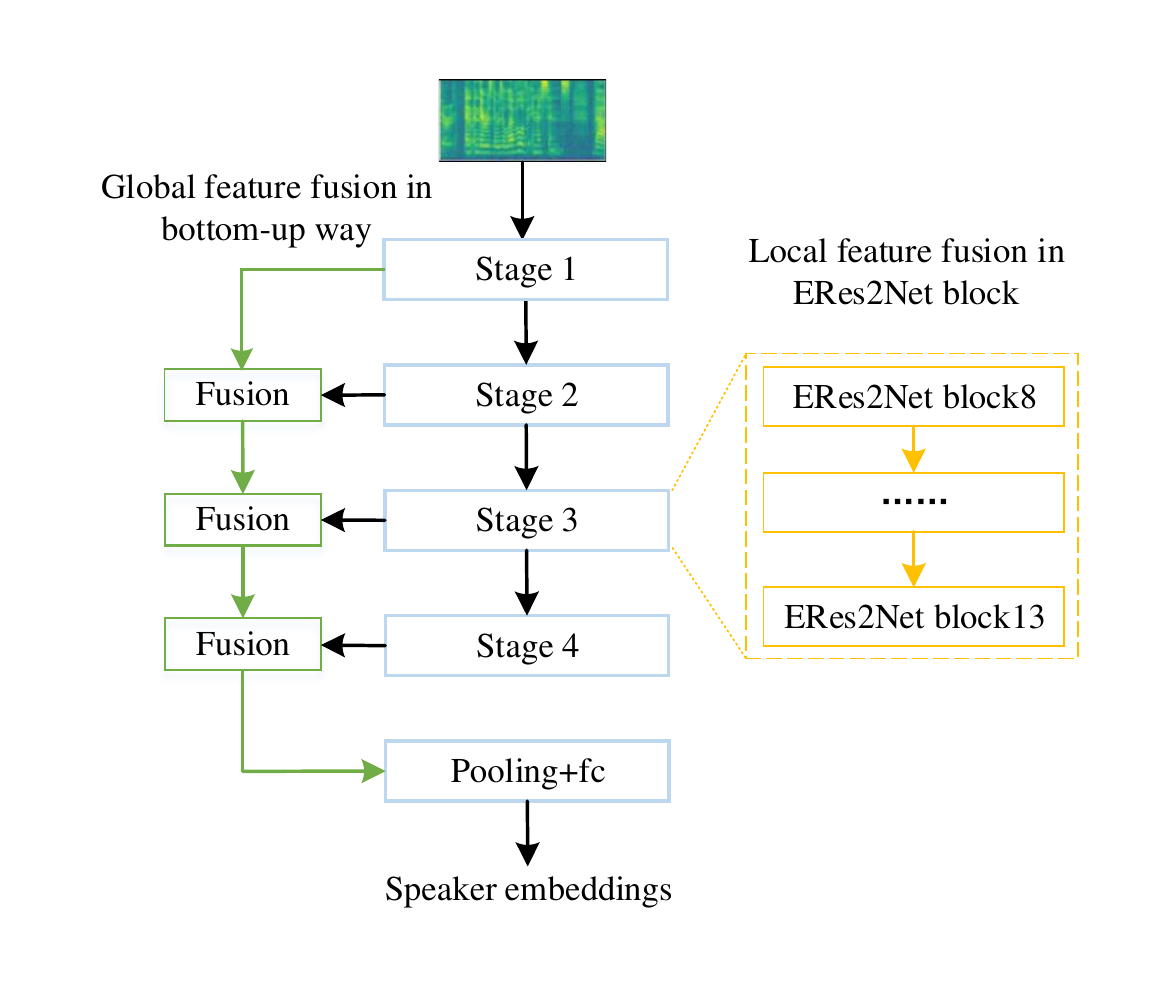}
  \caption{Overview of the enhanced Res2Net framework.}
\end{figure}

Although the architectures mentioned above are effective in speaker verification tasks, they may not fully capture the local and global information in the network. Additionally, the way of feature aggregation, such as element-wise addition or concatenation, provide a rigid combination of features that may not utilize the complementary information from different scales.
Inspired by the concept that human brain relies on a cognitive process with local and global information perception \cite{christie2012global}, we propose an Enhanced Res2Net architecture (ERes2Net) incorporates a local and global feature fusion mechanism for extracting speaker embeddings as shown in Fig. 1. The local feature fusion (LFF) component is designed to enhance the discrimination of speaker embeddings at a more granular level. It uses an attentional feature fusion (AFF) module to aggregate adjacent feature maps within residual block. And the global feature fusion (GFF) component modulates features of different temporal scales in bottom-up pathway and improves the robustness of speaker embedding from a global perspective.

We validate our proposed method on the VoxCeleb datasets. The experimental results show that ERes2Net architecture can achieve superior performance with fewer parameters. The remainder of this paper is organized as follows. In Section 2, we introduce the Res2Net and and its application in SV. In Section 3, we describe the ERes2Net with local and global feature fusion in detail. The experimental setup, the results and analysis are presented in Section 4. Finally, conclusions are given in Section 5.

\section{Res2Net for speaker verification}

\subsection{Res2Net block}
The Res2Net block\cite{DBLP:journals/pami/GaoCZZYT21} aims at improving model's multi-scale representation ability by increasing the number of available receptive fields. Within each residual block, it uses hierarchical residual-like connections to extract multi-scale features in the channel dimension. This approach achieves better performance in speaker verification especially for short utterances \cite{DBLP:conf/slt/ZhouZW21}. As shown in Fig. 2(a), feature maps were split into $s$ feature map subsets, denoted by $\textbf{x}_i$, where $i \in \{1,2,...,s\}$. Each feature subset $\textbf{x}_i$ has the same spatial size but $1/s$ number of channels. Every $\textbf{x}_i$ except $\textbf{x}_1$ goes through a $3\times3$ convolution filter $\textbf{K}_i()$. The output $\textbf{y}$ can be express as:
\begin{equation}
\mathbf{y}_i=\left\{\begin{array}{lr}
\mathbf{x}_i, & i=1 \\
\mathbf{K}_i\left(\mathbf{x}_i\right), & i=2 \\
\mathbf{K}_i\left(\mathbf{x}_i+\mathbf{y}_{i-1}\right), & i>2
\end{array}\right.
\end{equation}

\subsection{Speaker embedding network using Res2Net}

\begin{table}[thb]
\renewcommand\arraystretch{1.1}
\centering
\caption{The architecture of the speaker embedding network}
\scalebox{0.95}{
\begin{tabular}{cc|c|c|c}


\textbf{Stage} & \multicolumn{2}{|c|}{\textbf{Structure}} & \multicolumn{2}{c}{\textbf{Output size}}\\
\hline
\hline
 & \multicolumn{2}{|c|}{3 $\times$ 3, 32} &  \multicolumn{2}{c}{$T$ $\times$ 80 $\times$ 32} \\
\hline
\multirow{2}{*}[1ex]{Stage 1} & \multicolumn{2}{|c|}{$\begin{bmatrix} 1\times1 & 32 \\ \multicolumn{2}{c}{3\times3, 16, s = 2} \\ 1\times1 & 64 \\ \end{bmatrix}$ $\times$ 3} 
& \multicolumn{2}{c}{$T$ $\times$ 80 $\times$ 64} \\
\cline{1-5}
\multirow{2}{*}[1ex]{Stage 2} & \multicolumn{2}{|c|}{$\begin{bmatrix} 1\times1 & 64 \\ \multicolumn{2}{c}{3\times3, 32, s = 2} \\ 1\times1 & 128 \\ \end{bmatrix}$ $\times$ 4} 
& \multicolumn{2}{c}{$T/2$ $\times$ 40 $\times$ 128} \\
\cline{1-5}
\multirow{2}{*}[1ex]{Stage 3} & \multicolumn{2}{|c|}{$\begin{bmatrix} 1\times1 & 128 \\ \multicolumn{2}{c}{3\times3, 64, s = 2} \\ 1\times1 & 256 \\ \end{bmatrix}$ $\times$ 6} 
& \multicolumn{2}{c}{$T/4$ $\times$ 20 $\times$ 256} \\
\cline{1-5}
\multirow{2}{*}[1ex]{Stage 4} & \multicolumn{2}{|c|}{$\begin{bmatrix} 1\times1 & 256 \\ \multicolumn{2}{c}{3\times3, 128, s = 2} \\ 1\times1 & 512 \\ \end{bmatrix}$ $\times$ 3} 
& \multicolumn{2}{c}{$T/8$ $\times$ 10 $\times$ 512} \\
\cline{1-5}
\multirow{1}*{} & \multicolumn{4}{|c}{Temporal statistics pooling} \\
\cline{1-5}
\multirow{1}*{} & \multicolumn{4}{|c}{Fully connected layer} \\
\cline{1-5}
\multirow{1}*{} & \multicolumn{4}{|c}{Softmax layer} \\
\cline{1-5}
\end{tabular}}
\end{table}

A typical speaker embedding network contains (1) several frame-level layers obtaining high-level feature representations, (2) a segment-level pooling layer aggregating the frame-level representations across the temporal dimension and (3) a bottleneck layer projecting the pooled vector into a low-dimensional speaker embedding \cite{DBLP:conf/interspeech/SnyderGPK17}. We use Res2Net \cite{DBLP:conf/slt/ZhouZW21} as backbone to extract frame-level representations. Detailed configuration is listed in Table 1, where $T$ denotes variable-length data frames. The input layer consists of a single convolutional layer with a kernel size of 3$\times$3, stride of 1 and channel dimension of 32. Four residual stages include [3, 4, 6, 3] basic blocks with 64, 128, 256, and 512 channels output respectively. In each basic block, it has 3 convolutional layers with two filter sizes of 1$\times$1 and one filter size of 3$\times$3. After the 1$\times$1 convolution, feature maps were split into $s$ (we set $s$=2 in this work) feature map subsets. Every splitted feature goes through 3$\times$3 convolutional operator and concatenated. Then these feature maps are processed by an 1$\times$1 convolutional kernel. Down-sampling is performed by stage 2, stage 3 and stage 4 with a stride of 2$\times$2.

\begin{figure}[t]
  \centering
  \includegraphics[scale=0.63]{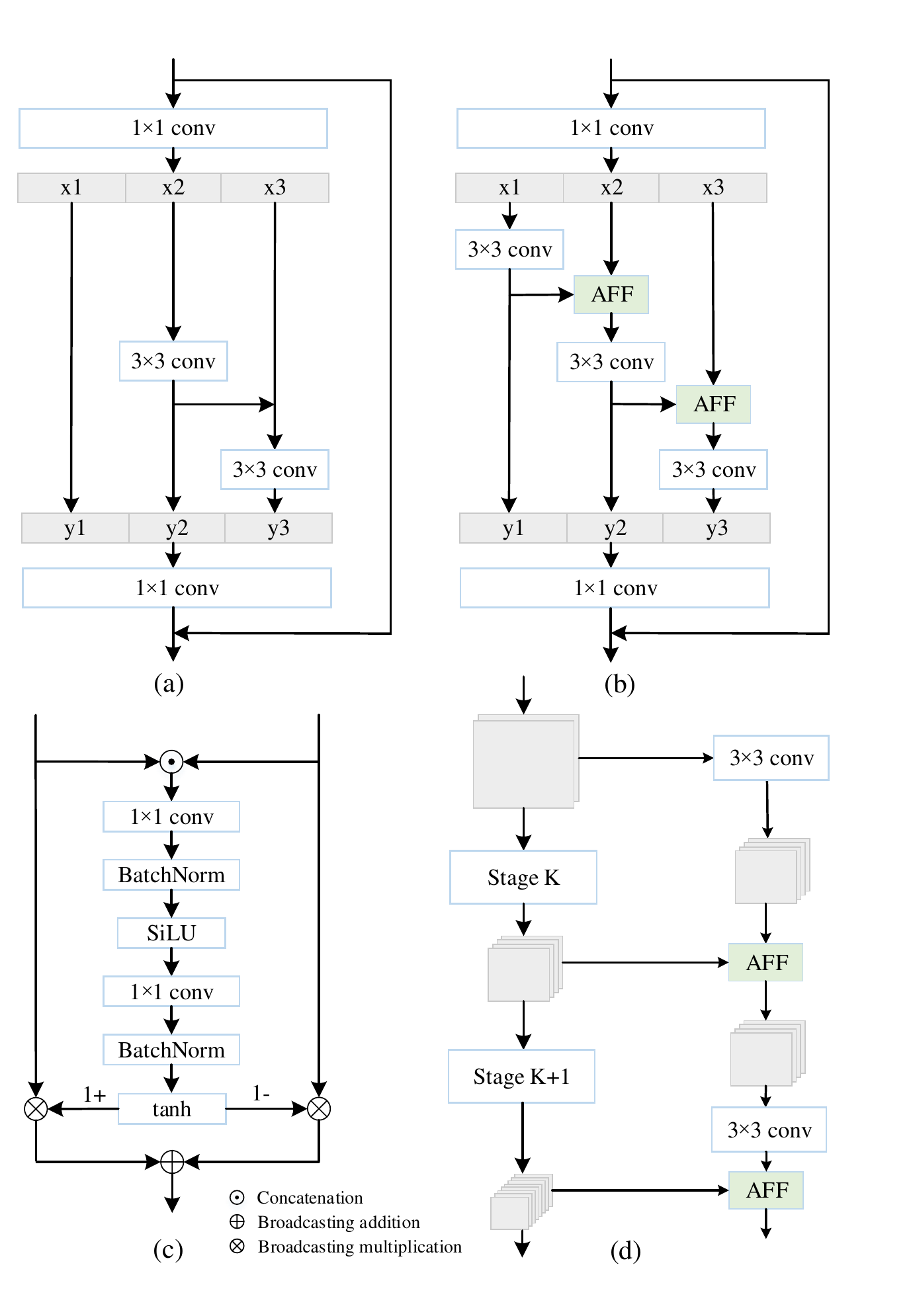}
  \caption{Illustration of different structures in the modules: (a) Res2Net block; (b) ERes2Net block; (c) Attentional feature fusion (AFF) module; (d) Global feature fusion (GFF) module.}
\end{figure}

\section{Proposed method}

\subsection{Overview of ERes2Net}
A popular concept is that human left hemisphere focus on local details while the right hemisphere pay more attention to global content \cite{christie2012global}. However, the split and concatenation strategy in Res2Net lacks effective local information interaction and global perspective from multiple temporal and frequency scales. To address this limitation, we propose an enhanced Res2Net architecture with local and global feature fusion.These techniques allow the ERes2Net architecture to capture both local and global patterns in the input signal, improving the accuracy and robustness of the speaker verification system. The overall architecture of the proposed network is depicted in Fig. 1 and consists of two branches: a local feature fusion branch and a bottom-up feature fusion branch. Fig. 2 dissects and analyzes each module of ERes2Net.

\subsection{Local feature fusion in ERes2Net}
This section describes the structure of proposed LFF in ERes2Net block, which introduces an attentional feature fusion mechanism in the residual-like connection between adjacent feature maps. The key idea of LFF is to obtain more fine-grained features and strengthen the local information interaction as shown in Fig. 2(b). 

In the ERes2Net block, feature maps $\mathbf{X}$ are divided into different groups according to the channel dimension after a $1 \times 1$ convolution, denoted by $\mathbf{x}_i$, where $i \in \{1,2,...,s\}$. We assume that $\mathbf{X} \in \mathbb{R}^{D \times T \times C}$ and each $\mathbf{x}_i \in \mathbb{R}^{D \times T \times C/s}$, where $D, T$ and $C$ denote the dimension of frequency, time and channel respectively. Then output features of the previous group are fused along with another group of input feature maps via an attentional feature fusion (AFF) module. AFF is used to strengthen the information interaction explicitly as shown in Fig. 2(c). Hierarchical fusion structure in the LFF block can increase the receptive fields of the model and integrate the local information across different channels. The output of ERes2Net block is derived as follows:
\begin{equation}
\mathbf{y}_{i}=\left\{\begin{array}{ll}
\mathbf{K}_{i}\left(\mathbf{x}_{i}\right), & i=1 \\
\mathbf{K}_{i}\left(\left(U\left(\mathbf{x}_{i}, \mathbf{y}_{i-1}\right)+1\right) \cdot \mathbf{x}_{i}+\right. & \\
\qquad \left.\left(1-U\left(\mathbf{x}_{i}, \mathbf{y}_{i-1}\right)\right) \cdot \mathbf{y}_{i-1}\right), & i>1
\end{array}\right.
\end{equation}

%
where $U(\cdot)$ represents the AFF module to compute local attention weights for adjacent feature maps. AFF module takes the concatenation of adjacent feature maps $\mathbf{x}_i$ and $\mathbf{y}_{i-1}$ as the input. Then calculate the local attention weights $\mathbf{U}$ as follows:
\begin{equation}
\mathbf{U} = tanh(\mathbf{BN}(\mathbf{W}_2 \cdot SiLU(\mathbf{BN}(\mathbf{W}_1 \cdot [\mathbf{x}_i, \mathbf{y}_{i-1}]))))
\end{equation}
where $[\cdot]$ denotes the concatenation along the channel dimension. $\mathbf{W}_1$ and $\mathbf{W}_2$ are point-wise convolution with output channel sizes of $C/r$ and $C$ respectively. $r$ is the channel reduction ratio (we set $r$=4 in this work). $\mathbf{BN}$ refers to batch normalization \cite{DBLP:conf/icml/IoffeS15}. $SiLU(\cdot)$ and $tanh(\cdot)$ stand for Sigmoid Linear Unit (SiLU) and tanh activation function respectively. This module is designed to dynamically weight and combine features based on their importance, improving the model's ability to extract relevant information from the input signal.

\subsection{Global feature fusion in ERes2Net}
This section describes the GFF component of the ERes2Net as shown in Fig. 2(d). The closer to the bottom layer of neural networks, the more limited the receptive field of neurons, and vice versa. GFF aims to enhance the global feature interaction by modulating features of different temporal scales in the bottom-up pathway. 

First of all, we select the multi-scale features $\{\mathbf{S}_j | j = 2,3,4\}$ of last layer in each ERes2Net stage which containing different temporal resolutions. Then we down-sample higher-resolution feature maps for ERes2Net stage output in the time and frequency dimension using $3\times3$ convolution kernel and expand the channel dimension with a factor of 2. Furthermore, we concatenate each stage output and calculate the bottom-up attention to modulate $\{\mathbf{S}_j | j = 2,3,4\}$ through AFF module. AFF module calculates the attention  weights from global perspective. The down-sampled feature maps are enhanced with features via AFF module as follows:
\begin{equation}
\mathbf{F}_j = U[\mathbf{D}(\mathbf{S}_{j-1}), \mathbf{S}_j] \qquad j = 2,3,4
\end{equation}
where $\mathbf{D}(\cdot)$ denotes the down-sampling operation. $\mathbf{F}_j$ stands for the fusion of $(j-1)-th$ stage output and $j-th$ stage output in bottom-up pathway.

\section{Experiments and analysis}

\subsection{Datasets and evaluation metrics}

To investigate the effectiveness of the proposed methods, we conduct experiments on the VoxCeleb datasets. The development portions of VoxCeleb2 \cite{DBLP:conf/interspeech/ChungNZ18} which comprises 1,092,009 utterances among 5,994 speakers are used for training. Performance of all systems are evaluated on the whole VoxCeleb1 \cite{DBLP:conf/interspeech/NagraniCZ17}. As shown in Table 2, there are three tasks on VoxCeleb1, and the last two tasks have more trials. Due to the background noise, reverberation and laughter contained in the speech data, three data augmentation techniques are applied to improve the robustness of the system: online data augmentation \cite{DBLP:journals/taslp/CaiCZL20} with MUSAN \cite{DBLP:journals/corr/SnyderCP15}, RIR dataset \cite{DBLP:conf/icassp/KoPPSK17}, and speed perturb \cite{wang2020dku} with 0.9 and 1.1 times speed changes to treble the number of speakers.

The results are reported in terms of two metrics, namely, the equal error rate (EER) and the minimum of the normalized detection cost function (MinDCF) with the settings of $P_{target}$ = 0.01 and $C_{fa} = C_{miss}$ = 1.
\vspace{-0.1cm}
\begin{table}[h]
    \caption{Tasks on VoxCeleb1 dataset. Here ‘O’ denotes ‘original’, ‘E’ denotes ‘extended’, and ‘H’ denotes ‘hard’}
    \scalebox{0.95}{
    \centering
        \begin{tabular}{cccccc}
        \toprule
         & VoxCeleb1-O & VoxCeleb1-E & VoxCeleb1-H \\
        \midrule
        Speakers & 40 & 1251 & 1251 \\
        Trials & 37,611 & 579,818 & 550,894 \\
        \bottomrule
        \end{tabular}}
\end{table}

\begin{table*}[thb]
    \caption{The experimental results on the VoxCeleb1-O, VoxCeleb1-E and VoxCeleb1-H evaluation protocols. The "Standard" refers to the standard configuration used in speaker verification (SV) systems as described in section 4.2, while "LMT" refers to another configuration described in the same section.}
	\centering
	\begin{tabular}                
        {c c p{1.2cm} p{1.2cm} p{1.2cm} p{1.2cm} p{1.2cm} p{1.2cm}}
	\toprule
        \multirow{2}*{\textbf{Architecture}} & \multirow{2}*{\textbf{Variant}} & \multicolumn{2}{c}{\textbf{VoxCeleb1-O}} & \multicolumn{2}{c}{\textbf{VoxCeleb1-E}} & \multicolumn{2}{c}{\textbf{VoxCeleb1-H}}\\
        \cmidrule(l){3-4} \cmidrule(l){5-6} \cmidrule(l){7-8}
		\cmidrule(l){3-4} \cmidrule(l){5-6} \cmidrule(l){7-8}
		 \multirow{1}*{} & \multirow{1}*{} & \textbf{EER(\%)} & \textbf{MinDCF} & \textbf{EER(\%)} & \textbf{MinDCF} & \textbf{EER(\%)} & \textbf{MinDCF} \\
        \midrule
        \midrule
        Res2Net & Standard & 1.51 & 0.148 & 1.38 & 0.148 & 2.40 & 0.224 \\
        Res2Net + LFF & Standard & 1.04 & 0.095 & 1.07 & 0.117 & 1.94 & 0.185 \\
        Res2Net + GFF & Standard & 1.33 & 0.132 & 1.28 & 0.138 & 2.33 & 0.217 \\
        ERes2Net & Standard & \textbf{0.92} & \textbf{0.094} & \textbf{0.99} & \textbf{0.111} & \textbf{1.92} & \textbf{0.181} \\
        \midrule
        \midrule  
        Res2Net & LMT & 1.44 & 0.120 & 1.26 & 0.135 & 2.18 & 0.207 \\
        Res2Net + LFF & LMT & 0.95 & 0.077 & 0.98 & 0.111 & 1.76 & 0.174 \\
        Res2Net + GFF & LMT & 1.26 & 0.108 & 1.19 & 0.129 & 2.09 & 0.199 \\
        ERes2Net & LMT & \textbf{0.83} & \textbf{0.072} & \textbf{0.94} & \textbf{0.104} & \textbf{1.79} & \textbf{0.173} \\
	 \bottomrule
    \end{tabular}
\end{table*}

\subsection{Systems configuration}

All of the systems used the Res2Net backbone as described in Table 1 and were trained using the PyTorch toolkit \cite{paszke2019pytorch}. We use the stochastic gradient descent (SGD) optimizer with a cosine annealing scheduler and a linear warm-up scheduler. During the first 5 epochs, the learning rate is linearly increased to 0.2. The momentum is set to 0.9 and weight decay to 1e-4. Angular additive margin softmax (AAM-Softmax) loss \cite{DBLP:journals/pami/DengGYXKZ22} is used for all experiments. The margin and scaling factors of AAM-Softmax loss are set to 0.3 and 32 respectively. The speaker embeddings are extracted from the first fully connected layer with a dimension of 192. Moreover, we use different training strategies depicted as follows.

\textbf{Standard}: The acoustic features used in the experiments are 80-dimensional Filter Bank (FBank) with 25ms windows and 10ms shift. To train the models, we randomly crop 3-second segments from each utterance for 70 epochs. Speech activity detection (SAD) is not performed as training data consists mostly of continuous speech. Mean and variance normalization is applied using instance normalization on FBank features. Cosine similarity is used for scoring.

\textbf{LMT}:  On the basis of standard configuration, we adopt the Large Margin Fine-tuning (LMT) \cite{DBLP:conf/icassp/ThienpondtDD21} strategy. Specifically, we abandon the speed perturbation augmentation and the speaker classification number is 5994. Besides, we randomly sample 6-second segments from each utterance to construct a training batch, and the margin in AAM is changed to 0.5. All the systems are trained for another 6 epochs. The initial value and final value of the learning rate are set to 1e-4 and 2.5e-5 respectively. Cosine similarity is used for scoring.

\subsection{Results and analysis}

We investigate the performance of proposed methods and evaluate them on the VoxCeleb1-O, VoxCeleb1-E and VoxCeleb1-H datasets. The experimental results are shown in Table 3. 

In the standard configuration systems, the proposed Res2Net with LFF achieves relative improvements in EER by 31.1\%, 22.5\% and 19.2\% when comparing rows 1 and 2, resulting in EER of 1.04\%, 1.07\% and 1.94\% on the three test sets, respectively. MinDCF has also been correspondingly improved.
The significant improvements achieved by the proposed Res2Net with LFF on all three test sets demonstrate the effectiveness of the local feature fusion mechanism in extracting speaker features. By emphasizing the fine-grained details in speaker characteristics, LFF enables the network to better capture the discriminative speaker embeddings.
Comparison between rows 1 and 3 shows that the proposed Res2Net with GFF achieves a relative EER reduction of 11.2\% and a relative MinDCF reduction of 10.8\% on the VoxCeleb1-O test set. This demonstrates the potency of fusing different temporal and frequency information for enhancing the robustness of speaker embeddings. 
However, it appears that LFF is more effective than GFF, possibly due to the AFF module's limited contribution in the global average. The combination of local feature fusion and global feature fusion in ERes2Net architecture yields superior performance, as demonstrated in row 4 of Table 3. It obtains 39.1\% , 28.3\% and 20\% relative EER reduction and 36.5\%, 25\% and 19.2\% relative MinDCF reduction on three test sets respectively. The experimental results indicate that the LFF and GFF methods are complementary, and highlight the necessity of combining local feature fusion and global feature fusion.

In the LMT configuration systems, we can draw the same conclusions as mentioned above. The ERes2Net system achieves the best performance, with EER of 0.83\% and MinDCF of 0.072 on the VoxCeleb1-O test set. Compared to the standard configuration, the performance has been significantly improved by applying the large margin fine-tuning strategy. The reason is that LMT strategy increases the inter-speaker distances between more reliable speaker centers while ensuring compact speaker classes. It can result in the generation of more speaker discriminative embeddings, provided that longer training utterances. Similar conclusions could be also found in \cite{DBLP:conf/icassp/ThienpondtDD21}.

\subsection{Comparison with published systems}
The proposed ERes2Net is compared with several state-of-the-art models with similar parameters. To ensure a fair comparison, we keep the training datasets the same with the reported systems. As presented in Table 4, we compared ERes2Net with other models such as D-TDNN, ResNet and Res2Net. Experimental results demonstrate that our model achieves better performance with fewer parameters, indicating the effectiveness and efficiency of our proposed approach for aggregating multi-scale features in time, frequency, and channel dimensions.

\begin{table}[h]
    \caption{Comparison with published systems in VoxCeleb1-O, * means that data augmentation is not used.}
    \scalebox{0.92}{
    \centering
    \begin{tabular}{p{3.1cm} c c c}
    \toprule
    \textbf{Framework} & \textbf{Params(M)} & \textbf{EER(\%)}  & \textbf{MinDCF} \\
    \midrule
    ResNet34-FPM* \cite{DBLP:conf/interspeech/JungKCJK20} & 5.85 & 1.98 & 0.205 \\
    Res2Net-MWA \cite{DBLP:journals/taslp/GuGZ23} & 5.50 & 1.71 & 0.165 \\
    Res2Net-26w8s \cite{DBLP:conf/slt/ZhouZW21} & 9.30 & 1.45 & 0.147 \\
    ResNet34-SKDFE* \cite{DBLP:conf/icassp/LiuWCWQ22} & 5.98 & 1.44 & 0.168 \\
    D-TDNN \cite{DBLP:conf/icassp/YuZSLL21} & 4.00 & 1.13 & 0.115 \\
    DF-ResNet56 \cite{DBLP:conf/interspeech/LiuCWWHQ22} & 4.49 & 0.96 & 0.103 \\
    \midrule
    \textbf{ERes2Net \ (Ours)} & \textbf{4.64} & \textbf{0.83} & \textbf{0.072} \\
    
    \bottomrule
    \end{tabular}}
\end{table}

\vspace{-0.35cm}
\section{Conclusions}

In this paper, we propose an enhanced Res2Net architecture with local and global feature fusion. ERes2Net is mainly composed of LFF and GFF. The LFF extracts localization-preserved speaker features and strengthen the local information interaction. GFF fuses multi-scale feature maps in bottom-up pathway to obtain global information. Both LFF and GFF employ the attentional feature fusion module proposed in this work. Experimental results on the VoxCeleb datasets show that the combination of local and global feature modeling can lead to robust speaker embedding extraction.

\bibliographystyle{IEEEtran}
\bibliography{mybib}

\end{document}